\pdfoutput=1
\documentclass[prl,twocolumn,superscriptaddress,preprintnumbers,nofootinbib]{revtex4}
\usepackage{graphicx}
\usepackage{epsfig}
\usepackage{bm}
\usepackage{latexsym,amssymb,amsmath,amsfonts,amssymb,txfonts,wasysym,float}
\usepackage{scrextend}
\usepackage{mathrsfs}
\usepackage{color}

\usepackage{enumitem}

\newcommand{\postscript}[2]{\setlength{\epsfxsize}{#2\hsize}
   \centerline{\epsfbox{#1}}}

\usepackage[usenames,dvipsnames]{xcolor}
\definecolor{orange}{cmyk}{0,0.5,1,0}
\definecolor{rossoCP3}{cmyk}{0,.88,.77,.40}
\definecolor{graa}{rgb}{0.8,0.8,0.8}
\definecolor{blaa}{rgb}{0.2,0.2,0.6}

\begin{document}
\preprint{MPP-2022-41}
\preprint{LMU-ASC 15/22}

\title{\color{rossoCP3} Anomalous $\bm{U(1)}$ Gauge Bosons and String Physics at the Forward Physics Facility}

\author{Luis A. Anchordoqui}

\affiliation{Department of Physics and Astronomy,  Lehman College, City University of
  New York, NY 10468, USA
}

\affiliation{Department of Physics,
 Graduate Center, City University
  of New York,  NY 10016, USA
}

\affiliation{Department of Astrophysics,
 American Museum of Natural History, NY
 10024, USA
}

\author{Ignatios Antoniadis}
\affiliation{Laboratoire de Physique Th\'eorique et Hautes \'Energies - LPTHE
Sorbonne Universit\'e, CNRS, 4 Place Jussieu, 75005 Paris, France
}

\affiliation{Nordita, Stockholm University and KTH Royal Institute of Technology
Hannes Alfv\'ens v\"ag 12, 106 91 Stockholm, Sweden}

\author{Karim Benakli}

\affiliation{Laboratoire de Physique Th\'eorique et Hautes \'Energies - LPTHE
Sorbonne Universit\'e, CNRS, 4 Place Jussieu, 75005 Paris, France
}

\author{Dieter\nolinebreak~L\"ust}

\affiliation{Max--Planck--Institut f\"ur Physik,  
 Werner--Heisenberg--Institut,
80805 M\"unchen, Germany
}

\affiliation{Arnold Sommerfeld Center for Theoretical Physics 
Ludwig-Maximilians-Universit\"at M\"unchen,
80333 M\"unchen, Germany
}

\begin{abstract}
  \vskip 2mm \noindent We show that experiments at the Forward Physics
  Facility, planned to operate near the ATLAS interaction point
  during the LHC high-luminosity era, will be able to probe predictions of Little String Theory by searching for anomalous $U(1)$ gauge bosons living in the
  bulk. The interaction of the abelian broken gauge symmetry with
the  Standard Model is generated at the one-loop level through
  kinetic mixing with the photon.
  Gauge invariant generation of mass for the $U(1)$ gauge boson proceeds
  via the Higgs mechanism in spontaneous symmetry breaking, or else
  through anomaly-cancellation appealing to St\"uckelberg-mass
  terms. We demonstrate that FASER2 will be able to probe string
  scales over roughly two orders of magnitude: $10^5 \alt M_s/{\rm TeV} \alt 10^7$.
\end{abstract}
\maketitle

In Ref.~\cite{Anchordoqui:2020tlp} we investigated the sensitivity of
dark matter direct detection experiments to extremely weakly coupled
extra $U(1)$ gauge symmetries which are ubiquitous in D-brane string
compactifications~\cite{Blumenhagen:2005mu,Blumenhagen:2006ci}. In
this follow up to the dark matter work we particularize our
investigation to experiments planned to operate at the HL-LHC Forward
Physics Facility (FPF)~\cite{Anchordoqui:2021ghd,Feng:2022inv}. Before
proceeding, we pause to stress that our investigation will be framed within
the context of Little String Theory (LST), which allows us to take
the string coupling $g_s$ of arbitrary small values~\cite{Antoniadis:2001sw,Antoniadis:2011qw}. This contrasts with previous
literature on hidden $U(1)$ in string theory, which pivots on the
volume of the internal space rather than on $g_s$.

\begin{figure}[t]
      \postscript{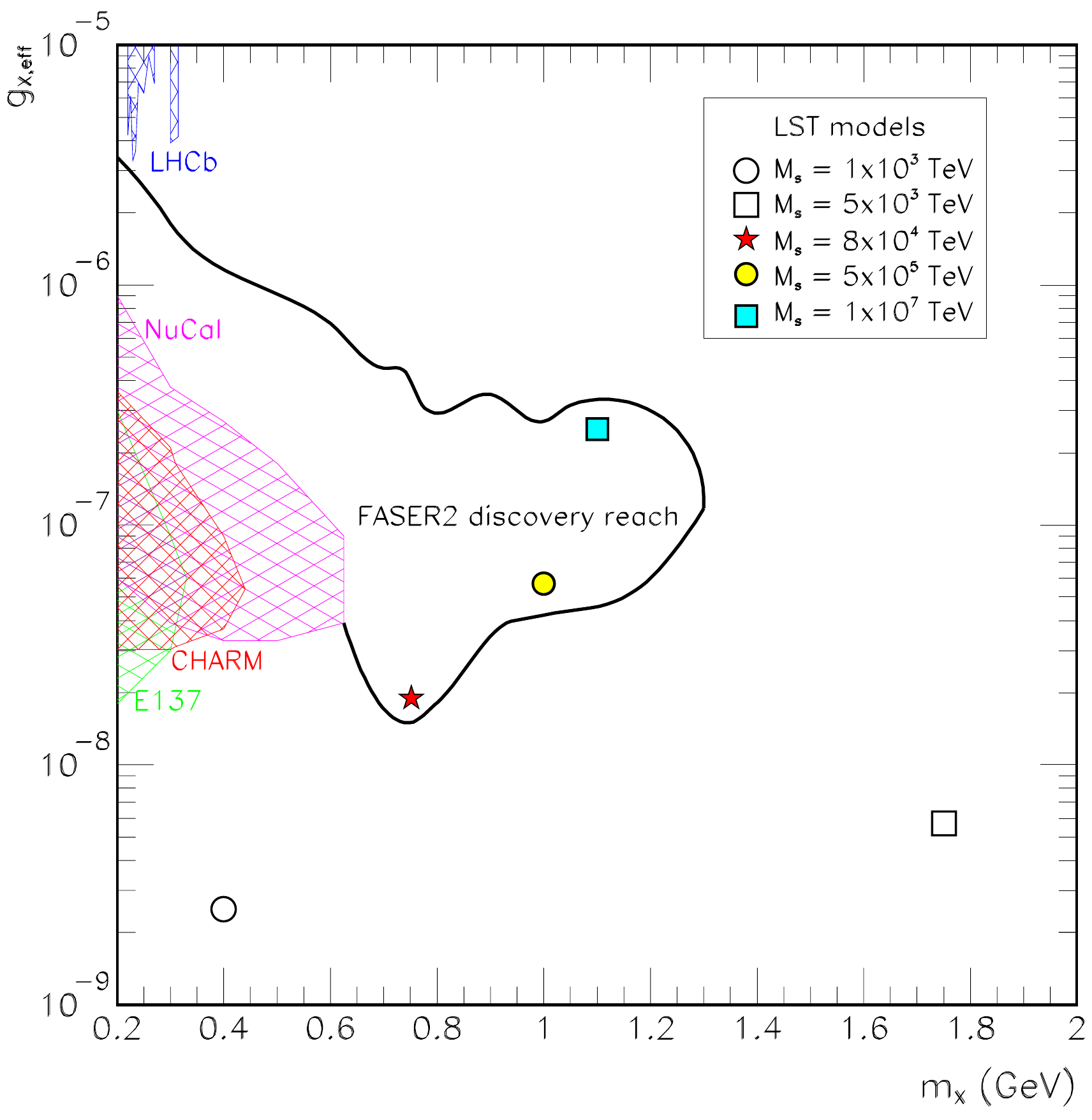}{0.91}
      \caption{Dark $U(1)_X$ sensitivity reach contour in the
        $(m_X, g_{X,{\rm eff}})$ plane obtained with the {\tt FORSEE}
        package~\cite{Kling:2021fwx}. The shaded regions are excluded by previous
        experiments: SLAC (E137)~\cite{Bjorken:1988as}, CHARM~\cite{Gninenko:2012eq}, NuCal~\cite{Tsai:2019buq}, and LHCb~\cite{LHCb:2017trq}. The symbols indicate particular predictions
        of LST models for feasible 
        parameters given in Table~\ref{tabla1}. \label{fig:1}}
      \end{figure}

We focus attention on the FPF's second generation ForwArd Search ExpeRiment
(FASER2).\footnote{FASER has been already installed in the LHC tunnel and
  will collect data during Run 3~\cite{Feng:2017uoz}.} FASER2 will be
shielded from the ATLAS interaction point by 200~m of concrete and
rock, creating an extremely low-background environment for searches of
long-lived particles traveling unscathed along the beam collision
axis. Herein, we are interested in searches for light, very
weakly-interacting vector fields that couple through kinetic mixing to
the hypercharge gauge boson or, at low energies, effectively to the
Standard Model photon (SM). At hadron colliders like the LHC, dark
$U(1)_X$ gauge bosons of mass $m_X$ can be abundantly produced through
proton bremsstrahlung or via the decay of heavy mesons. Indeed, over
the lifetime of the HL-LHC there will be
 $4 \times 10^{17}$ neutral pions, $6\times 10^{16} \, \eta$ mesons,
 $2 \times 10^{15} \, D$ mesons, and $10^{13} \, B$ mesons produced in the
 direction of FASER2. The $U(1)_X$ discovery potential of FASER2 in the $(m_X,g_{X,{\rm eff}})$ plane is
  shown in Fig.~\ref{fig:1}, where we have defined the effective kinetic mixing parameter $g_{X,{\rm eff}} \equiv e
    \epsilon_{\gamma X}$, and where $e$ is the elementary charge and
    $\epsilon_{\gamma X}$ is the physical kinetic mixing
    parameter. We note in passing that complementary measurements of
    weakly coupled $U(1)$ gauge bosons could 
    be carried out by the proposed CERN  experiments SHiP~\cite{Alekhin:2015byh},  FACET~\cite{Cerci:2021nlb}, and MATHUSLA~\cite{MATHUSLA:2019qpy}.

We now turn to demonstrate that LST provides a compelling framework
for engineering very weak extra gauge
symmetries with masses $500 < m_X/{\rm MeV} < 800$ and $10^{-8} < g_{X,{\rm
    eff}} < 10^{-7}$. The SM gauge group
is localized on Neuveu-Schwarz (NS) branes (dual to the
D-branes). However, the
$U(1)_X$ gauge field could live in the bulk and if so its four-dimensional gauge coupling becomes infinitesimally small~\cite{Antoniadis:2002cs}.

We consider a compactification on a six-dimensional space with the Planck mass given by
\begin{equation}
 M_{\rm Pl}^2 = \frac{8}{g_s^2} \ M_s^8 \ \frac{V_2 V_4}{(2 \pi)^6} 
\label{mpl}
\end{equation}
(up to a factor 2 in the absence of an orientifold), where $M_s$ is
the string scale and we  have taken the
internal space to be a product of a two-dimensional space, of volume
$V_2$, times a four-dimensional compact space, of volume $V_4$. We
further consider that the SM degrees of freedom emerge on a stack of
NS5-branes wrapping the two-cycle of volume $V_2$. For simplicity, we
assume that the internal space is characterized by a torus made of two orthogonal circles with radii $R_1$ and $R_2$. The corresponding (tree-level) gauge coupling is given by:
\begin{equation}
g_{\rm SM}^2 = \frac{R_1}{R_2} \quad ({\rm Type \, \,  IIA}) \quad {\rm and}
\quad g_{\rm SM}^2 = \frac{1}{R_1 R_2 M_s^2}  \quad ({\rm Type \,  \,  IIB}); 
\label{gSM}
\end{equation}
hence, an order one SM coupling imposes $R_1\simeq  R_2 \simeq
M_s^{-1 }$. The $U(1)_X$ gauge field lives on a  D$(3+\delta_X)$-brane that wraps a $\delta_X$-cycle of volume $V_X$, while its remaining
four dimensions extend into the uncompactified space-time. The corresponding gauge coupling is given by,
\begin{equation}
    g_X^2 =\frac{(2 \pi)^{\delta_X +1} \ g_s}{V_X \ M_s^{\delta_X}} \,,
\label{gb}
  \end{equation} 
If $U(1)_X$ arises from a D7-brane (\ref{gb}) can be recast as 
\begin{equation}
g_X^2 =\frac{(2 \pi)^{5} \ g_s}{V_4 \ M_s^4} \,.
\label{gx}
\end{equation}
We further assume that all the internal space radii are of the order of the
string length, $M_s^6 V_2 V_4 \simeq (2 \pi)^6$, so that (\ref{mpl}) and (\ref{gx}) yield,
\begin{equation}
g_{X}\simeq \ 32^{1/4} \sqrt{\pi} \ \sqrt{\frac{M_s}{M_{\rm Pl}}} \, .
\label{LST-coupling-X}
\end{equation}

The dark gauge boson acquires a mass through the Higgs mechanism, with $v_X$ the vacuum
expectation value for the Higgs $h_X$ that breaks the $U(1)_X$
symmetry. We consider the simplest quartic potential $-\mu_X^2 h_X^2 +
\lambda_X h_X^4$, which gives $v_X = {\mu_X}/\sqrt{{2 \lambda_X}}$, a
Higgs mass of order $\mu_X$, and a mass for the  dark gauge boson 
\begin{equation}
m_X =  \frac{g_X  \mu_X}{\sqrt{{2 \lambda_X}}} =   \sqrt{2 \pi g_s} \left(
  \frac{8}{g_s^2} \right)^{\delta_X/2 d} \  \left(\frac{M_s}{M_{\rm Pl}}
\right)^{\delta_X/d} \, \,  v_X \, ,
\label{mass-Higgs-delta}
\end{equation}
where we have taken $\lambda_X$ to be ${\cal O} (1)$, and where $d$ is
the total number of dimensions that are large. Throughout  our calculations we take $d=\delta_X$.

Alternatively, the abelian gauge field $U(1)_X$ could acquire a mass via a St\"uckelberg mechanism as a consequence
of a Green-Schwarz (GS) anomaly cancellation~\cite{Green:1984sg,Dine:1987bq}, which is achieved through the coupling of
twisted Ramond-Ramond axions~\cite{Sagnotti:1992qw, Ibanez:1998qp}. The mass of the
anomalous $U(1)_X$ can be unambiguously
calculated through a direct one-loop string computation, and is given by
\begin{equation}
  m_X = \varkappa \ \sqrt{\frac{V_a M_s^{2}}{V_X M_s^{\delta}}} M_s =  \frac{\varkappa}{(2 \pi)^{\frac{\delta_x -2}{2}}} 
  \left( \frac{\sqrt{8}}{g_s} \frac{M_s}{M_{\rm Pl}} \right) ^{\frac{\delta_X- \delta_a}{d}} M_s\,, 
\label{Stuckelberg}
\end{equation}
where $\varkappa$ is the GS anomaly coefficient (which is a numerical
factor of order $10^{-1}$ to $10^{-2}$ times $\sqrt{g_s} \propto g_X$), $V_a$ is the two-dimensional
internal volume corresponding to the propagation of the axion
field~\cite{Antoniadis:2002cs} and $ \delta_a$ is the number of large
dimensions in $V_a$.\footnote{Note
  that the $U(1)$ is not necessarily anomalous in four dimensions. A
  mass can be generated for a non-anomalous $U(1)$ by a
  six-dimensional (6d) GS term associated to a 6d anomaly cancellation
  in a sector of the theory.} To develop some sense for the orders of
magnitude involved, in our calculations we take $\delta_a =2$ and $ \delta_X =d=4$. With this in mind, (\ref{Stuckelberg}) can be
rewritten as 
\begin{equation}
  m_X =  \frac{\varkappa}{(2 \pi)} 
  \left( \frac{\sqrt{8}}{g_s} \frac{M_s}{M_{\rm Pl}} \right)
  ^{\frac{1}{2}} M_s \, .
  \label{Stuckelberg2}
\end{equation}
For a concrete example of this set up, we envision 2
D7-branes intersecting in two common directions; {\it viz.}, $D7_1\!:1234$
and $D7_2\!:1256$, where $123456$ denote the internal six
directions. Next, we take 1234 to be large and 56 to be small (i.e, order the string scale)
compact dimensions. The gauge fields of $D7_1$ have a suppression of
their coupling by the 4-dimensional internal volume $V_X$ while the
states in the intersection of the two D7 branes only see the 12 large
dimensions and lead to 6 dimensional anomalies, which are cancelled by an axion
living in the same intersection, so $V_a$ is the volume of 12
only. 

As noted above, the $U(1)_X$ does not couple directly to the visible
sector, but does it via kinetic mixing with ordinary photons. This coupling can be generated by non-renormalizable operators, but it is natural to assume that it is generated by loops of states carrying charges $(q^{(i)}, q_X^{(i)})$ under the two $U(1)$'s and having masses $m_i$:
\begin{equation}
\epsilon_{\gamma X} =\frac{e g_X  }{16 \pi^2} \sum_i q^{(i)}
q_X^{(i)} \ln {\frac{ m_i^2}{{\mu^2}}} \equiv \frac{e g_X }{16 \pi^2}
C_{\rm Log} 
\end{equation}
where $\mu^2$ denotes the renormalization scale (which in string
theory is replaced by $M_s$), and where we absorbed also the constant
contribution. The effective coupling to SM is then given by 
\begin{equation}
g_{X, {\rm eff}} = e \epsilon_{\gamma X}=   \frac {\alpha_{\rm em} g_X }{4 \pi}
C_{\rm Log} \, .
\label{gXeff}
\end{equation}

Using (\ref{LST-coupling-X}) and (\ref{mass-Higgs-delta}), as well as
(\ref{Stuckelberg2}) or (\ref{gXeff}) we scan over the LST parameter
space. Our
results are encapsulated in Fig.~\ref{fig:1}, where we show representative
values of the $(m_X, g_{X,{\rm eff}})$ plane, with LST model
parameters listed in Table~\ref{tabla1}.
In our calculations we set $C_{\rm Log} \sim
3$~\cite{Rizzo:2018vlb}. A point worth noting at this juncture is that
for $g_{X,{\rm eff}} \agt 5 \times 10^{-8}$, the associated abelian gauge
boson can only acquire a mass via the Higgs mechanism, since the required GS numerical factor becomes unnaturally small. We can explicitly see in
  the figure that the LST parameter
  space region probable by FASER2 spans roughly two orders of
  magnitude in the string scale.

\begin{table}
\caption{Selected LST model parameters of points shown in
  Fig.~\ref{fig:1}. \label{tabla1}}
  \begin{tabular}{cccccc}
    \hline\hline
    Symbol & $M_s$ (TeV) & $g_s$ & $v_x/M_s$ & $\varkappa$ \\ \hline
    ~~$\circ$~~ &  ~~$1 \times 10^3$~~ & ~~$2.4 \times 10^{-13}$~~ & ~~$3.4 \times
                                                         10^{-1}$~~   &
                                                                      ~~$2.6
                                                                      \times
    10^{-6}$~~ \\
    $\square$ & $5 \times 10^3$ &$1.2 \times 10^{-12}$ & $1.3 \times
                                                         10^{-1}$  &
                                                                     $2.3
                                                                     \times
    10^{-6}$\\
    $\star$ &  $ 8 \times 10^4$ & $1.9 \times 10^{-11}$ & $8.7 \times
                                                          10^{-4}$ & $5.8
                                                                  \times
    10^{-8}$ \\
    $\bullet$ & $ 5 \times 10^5$ & $1.2 \times 10^{-10}$ & $7.5 \times
                                             10^{-5}$ & $1.3 \times
                                                        10^{-8}$ \\
    $\blacksquare$ & $ 1 \times 10^7$ & $2.4 \times
                                        10^{-9\phantom{0}}$ & $9.4
                                                              \times 10^{-7}$
                                             & $-$ \\
    \hline \hline
    \end{tabular}
\end{table}

In summary, we have shown that FASER2 will be able to probe a region
of the LST parameter space by searching for abelian gauge bosons living in
the bulk. From (\ref{mass-Higgs-delta}) we see that the effective coupling $g_{X,{\rm eff}}$
depends on the product $\sqrt{M_s} \times C_{\rm Log}$. Similar ${\cal O} (1)$
logarithmic terms appear when computing threshold corrections to gauge
couplings. For $C_{\rm Log}$ of order one, we can see in
Fig.~\ref{fig:1} that there is a minimum value of the string scale,
roughly of order $10^5~{\rm TeV}$, which is in the FASER2 probable
region. There is also a maximum value that can be tested in this
region, $M_s \alt 10^7~{\rm TeV}$. For the mass of the hidden Higgs,
one needs some hierarchy given by $v_X/M_s$ as shown in
Table~\ref{tabla1}. Given these considerations, the mass of the hidden
$U(1)_X$ scales as  $(v_x/M_s) \times M_s^{3/2}$, so the hierarchy
increases as $M_s^{3/2}$ for fixed $m_X$.

In closing, we note that the SM singlet scalar field $S$ would couple to the SM Higgs doublet
$H$, yielding a portal into the dark
sector~\cite{Patt:2006fw}. However, for $\mu_X \gg m_H$, the $S \leftrightharpoons H$ mixing angle $ \ll 1$, where $m_H$ is the Higss mass~\cite{Weinberg:2013kea}. Using the values of
$v_X$ given in Table~\ref{tabla1}, it is straightforward to see that the Higgs
portal does not generate additional bounds on the model and that the
hidden scalar is out of the LHC reach; e.g., for $M_s \sim 8 \times
10^4~{\rm TeV}$, we have $\mu_X \sim 100~{\rm TeV}$.\\

The work of L.A.A. is supported by the U.S. National Science
Foundation (NSF Grant PHY-2112527). The work of K.B. is supported by the Agence
Nationale de Recherche under grant ANR-15-CE31-0002
``HiggsAutomator''. The work of D.L. is supported by the Origins
Excellence Cluster.

\end{document}